\documentclass[aps,prc,twocolumn,floatfix,superscriptaddress,nofootinbib,preprintnumbers,longbibliography,amsmath,amsthm,amssymb]{revtex4-2}
\usepackage{graphicx}
\usepackage{amsfonts}
\usepackage{color}
\usepackage[colorlinks=true,linkcolor=blue,citecolor=blue]{hyperref}
\usepackage{dcolumn}
\usepackage{bm}
\usepackage{braket}

\newcommand{\br}{\boldsymbol{r}}

\begin{document}
\allowdisplaybreaks[1]
\title{
Proton--neutron pair correlations in neutron-rich nuclei
}
\author{Kenichi Yoshida}
\email[E-mail: ]{kyoshida@rcnp.osaka-u.ac.jp}
\affiliation{Research Center for Nuclear Physics, Osaka University, 
Ibaraki, Osaka 567-0047 Japan}
\affiliation{RIKEN Nishina Center for Accelerator-Based Science, Wako, Saitama 351-0198, Japan}
\affiliation{Center for Computational Sciences, University of Tsukuba, Tsukuba, Ibaraki 305-8577, Japan}
\date{\today}
\begin{abstract}
\begin{description}
\item[Background]
Nuclear pairing is a well-established many-body correlation, particularly among like particles in a spin-singlet state. However, the strength of spin-triplet proton--neutron (pn) pairing in nuclei has remained 
a long-standing and unresolved issue.
\item[Purpose]
The relative strength of spin-triplet pn pairing 
compared to spin-singlet one is investigated 
by introducing and analyzing the polarizability of the response to pn pair transfers.
\item[Method] 
The nuclear energy-density functional method is employed. 
The ground state of the target nucleus is described 
using the Hartree--Fock--Bogoliubov approximation, 
which accounts for the conventional superfluidity of like-particle pairs. 
The response to pn pair transfers is then analyzed using the 
pn quasiparticle random-phase approximation.
\item[Results]
The spin-singlet pn-pair correlation is strongest at $N=Z$ 
and decreases monotonically with the increasing number of excess neutrons, 
whereas the spin-triplet pn-pair correlation is 
shown to depend non-monotonically on the neutron number 
and can be enhanced in cases where the 
pn-pair transfers involving the $\pi j_> \otimes \nu j_<$ 
configuration occur at low energy.
\item[Conclusions]
The shell effect, which uniquely appears in spin-triplet pn-pair correlation, 
serves as a key indicator of the strength of pn-pair correlations.
\end{description}

\end{abstract}
\maketitle
\section{Introduction}\label{intro}
Pairing is a correlation occurring universally in
assemblages made of interacting many particles, 
ranging from quark and nuclear to electron many-body systems. 
In infinite systems, the correlation is quantified by
the introduction of an order parameter, specifically the pairing gap, which arises from condensation.
When the correlation is sufficiently strong, fluctuations become significant, leading to a phase transition and the emergence of condensation.

An analogy to superconductivity has been applied 
to finite nuclei by employing 
the Bardeen--Cooper--Schrieffer (BCS)
theory~\cite{PhysRev.106.162} to the nuclear Hamiltonian~\cite{BM2,rin80}. 
Closed-shell nuclei are in a normal phase, 
where the gap function vanishes, 
while most nuclei exhibit a superconducting/superfluid 
phase at low energy. 
A non-vanishing pairing gap in the spin-singlet channel 
plays a central role in various low-energy nuclear phenomena, 
including the ground-state spin, 
staggering in the systematics of binding energies, 
low-lying quadrupole collective dynamics, 
rotational moments of inertia, 
and spontaneous fission half-lives~\cite{doi:10.1142/9789814412490_0003}. 
A notable feature unique to finite systems is that pairing can be probed in response to a field changing 
the number of particles; 
the pair correlation leads to an enhanced 
two-nucleon transfer reaction~\cite{YOSHIDA1962685}.

Heavy proton-rich nuclei along the $N=Z$ line 
have received a great deal of attention 
with the advancement of experimental techniques 
in rare-isotope beams for nuclear physics. 
Of particular interest is the possible appearance of 
exotic phenomena, 
attributed to the coherent shell effects of both protons and neutrons.  
Specifically, the spin-triplet proton--neutron (pn) 
pair correlation 
is expected to be visible, 
as the spatial overlap between the proton and neutron 
single-particle wave functions would be large to 
form a Cooper pair of a proton and a neutron---a phenomenon 
initially predicted in light $N=Z$ nuclei~\cite{PhysRev.140.B26}. 
Moreover,  
the possible coexistence of the spin-singlet 
pn-pair condensation with the spin-triplet one 
is predicted~\cite{Goodman:1999zz, Gezerlis:2011rh}.

Spin-orbit splitting is crucial in determining 
whether spin-triplet or spin-singlet pair correlation is stronger in medium-heavy nuclei~\cite{POVES1998203}. 
If a nucleus is large enough that the surface spin-orbit potential can be neglected, 
the spin-triplet pn-pair superfluid appears 
in the ground state~\cite{Bertsch:2009xz}, 
as is expected in symmetric nuclear matter~\cite{alm90, VONDERFECHT19911, BALDO19928, 10.1143/PTP.112.27}.
However, in reality, spin-orbit splitting suppresses 
spin-triplet pn-pair correlation in finite nuclei. 
Consequently, despite numerous attempts, 
the experimental evidence for spin-triplet 
pair condensation remains a topic of ongoing debate~\cite{Frauendorf:2014mja}.

The present work further investigates the role of 
spin-orbit splitting 
to garner a deeper insight into 
the pn-pair correlations in medium-heavy nuclei. 
Previous studies have focused primarily on 
$N \approx Z$ nuclei, 
where the correlation 
between a proton and a neutron is the strongest. 
However, as demonstrated in Ref.~\cite{Gezerlis:2011rh}, 
stepping away from the $N=Z$ line 
can also reveal unexpected and novel findings. 
In this study, 
I explore the pn-pair correlations across a wide mass region 
from calcium to tin isotopes, 
extending from the $N=Z$ line to the neutron-rich side. 
To systematically study the pn-pair correlations, 
I employ the notion of the fluctuation of order parameters.

The paper is organized as follows. 
In Sec.~\ref{Sec.Th}, 
I rehash the static polarizability/susceptibility 
in response to an external field 
and then extend it to the case for 
the response to pn-pair transfers. 
A numerical method I employ is briefly summarized 
in Sec.~\ref{Sec.numer}. 
In Sec.~\ref{Sec.Result}, 
I show the results and discuss the roles of excess neutrons 
on the pn-pair polarizabilities.
Section \ref{Sec.Sum} summarizes the paper.

\section{Theoretical framework}\label{Sec.Th}
\subsection{Static polarizability}\label{Sec.pol}
To explore the collectivity, 
one may perform the constrained Hartree--Fock--Bogoliubov
(HFB) calculations in the presence of 
an external field $\lambda \hat{F}$ relevant to the collective mode, i.e. 
one solves 
\begin{align}
    \delta \bra{\phi(\lambda)}\hat{H}-\lambda \hat{F}\ket{\phi(\lambda)}=0,
\end{align}
where $\hat{H}$ is the Hamiltonian and $\ket{\phi}$ the HFB state vector.
The static polarizability $\alpha$ is defined by the change of 
the expectation value of $\hat{F}$~\cite{rin80}:
\begin{align}
    \bra{\phi(\lambda)}\hat{F}\ket{\phi(\lambda)}=\bra{\phi_0}\hat{F}\ket{\phi_0}+\lambda \alpha
\end{align}
and the change in energy is given by
\begin{align}
    \bra{\phi(\lambda)}\hat{H}\ket{\phi(\lambda)}=\bra{\phi_0}\hat{H}\ket{\phi_0}+\dfrac{1}{2}\lambda^2 \alpha,
\end{align}
where $\ket{\phi_0}=\ket{\phi(0)}$ is the solution of the HFB equation.
One then evaluates the static polarizability, 
the radius of curvature of the potential energy surface 
at the HFB equilibrium, 
as
\begin{align}
    \alpha=\dfrac{d^2}{d\lambda^2}\bra{\phi(\lambda)}\hat{H}\ket{\phi(\lambda)}|_{\lambda=0}.
\end{align}

Alternatively, one can compute the response functions $R_F(\omega)$ 
of the system to the external field $\hat{F}$ in the random-phase 
approximation (RPA):
\begin{align}
    R_F(\omega)=\sum_n\left(\dfrac{|\bra{n} \hat{F} \ket{0}|^2}{\omega-\omega_n+i\eta }
    -\dfrac{|\bra{0} \hat{F} \ket{n}|^2}{\omega+\omega_n+i\eta }\right),
\end{align}
where $\ket{n}$ and $\omega_n$ denote the RPA eigenstates and eigenenergies. 
If one is interested in the static polarizability, 
it is sufficient to calculate the inversely-energy weighted sum (IEWS) of the strength distribution:
\begin{align}
    \alpha=-R_F(0)=2m_{-1}(F),
\end{align}
with $m_k(F)$ being the $k$-th moment of the strength distribution. 
This is nothing but the dielectric theorem~\cite{Capelli:2009zz}.

It is noted that the IEWS and the static polarizability have been
discussed intensively for the case of the $E1$ strength function, i.e. the electric dipole excitation
in nuclei, 
as a probe of neutron skin and nuclear matter properties~\cite{PhysRevLett.107.062502, Piekarewicz:2012pp,Roca-Maza:2018ujj}. 
Quite recently, the pair collectivity has been 
studied in terms of the pair polarizability~\cite{Takahashi:2023rfn}.

\subsection{Polalizability for proton--neutron pairing}\label{Sec.pn_pol}
The pair field with spin $S$ and isospin $T$ is defined as
\begin{align}
\hat{P}_{ST}(\boldsymbol{r}) :=&
\dfrac{1}{2\sqrt{2}}\sum_{s s'}\sum_{t t'} 
\hat{\psi}(\boldsymbol{r}\tilde{s}' \tilde{t}')
\hat{\psi}(\boldsymbol{r}s t)
\mathsf{S}_{S \mu_\sigma, s' s}
\mathsf{T}_{T \mu_\tau, t' t}
\label{pair_field} 
\end{align}
with the nucleon field operator $\hat{\psi}(\br s t)$, 
and 
$\hat{\psi}(\boldsymbol{r} \tilde{s} \tilde{t}) 
:= (-2s)(-2t)\hat{\psi}(\boldsymbol{r}-s-t)$. 
The spin and isospin operators are defined as
\begin{align}
    &\mathsf{S}_{00}=1, \hspace{0.5cm} \mathsf{S}_{1\mu_\sigma}=\sigma_{\mu_\sigma}, \\
    &\mathsf{T}_{00}=1, \hspace{0.5cm} \mathsf{T}_{1\mu_\tau}=\tau_{\mu_\tau}.
\end{align}
Here, 
$\sigma_{\mu_\sigma}$ and $\tau_{\mu_\tau}$ 
are the spherical components of 
spin and isospin Pauli matrices, respectively; 
$s=\pm 1/2$ stands for the $z$-component of spin; 
$t=\pm 1/2$ stands for neutron and proton. 
The factor of $1/\sqrt{2}$ is put here so that 
the two-neutron pair field ($\mu_\tau=1$) 
coincides with the pair field operator given in Ref.~\cite{Takahashi:2023rfn}. 
Note that it is convenient to introduce the factor 
of $1/\sqrt{2}$ 
when investigating the isovector modes of excitation  
with $T_z=\pm1$ and $T_z=0$ simultaneously~\cite{Yoshida:2021gla}.

One can expect the pn-pair correlations to be probed by the response 
of the pn-pair transfers~\cite{FROBRICH1971338, PhysRevLett.94.162502}. 
I then introduce an $L=0$ pn-pair addition operator 
\begin{align}
    \hat{P}^{{\rm ad}}_{ST}
    :=\int d\boldsymbol{r}Y_{00}(\hat{r})f(r)\hat{P}^\dagger_{ST}(\boldsymbol{r})
    \label{eq:pair_ad}
\end{align}
and an $L=0$ pn-pair removal operator
\begin{align}
    \hat{P}^{{\rm rm}}_{ST}
    :=\int d\boldsymbol{r}Y_{00}(\hat{r})f(r)\hat{P}_{ST}(\boldsymbol{r}),
    \label{eq:pair_rem}
\end{align}
with $\mu_\tau=0$ for $T=1$. 
These operators bring about a transition, 
which changes the particle number $\Delta N= \pm 2$ 
with $\Delta T_z=0$. 
As in the study of a two-neutron transfer~\cite{Takahashi:2023rfn}, 
I introduced a form factor $f(r)$, 
which is effective in a spatial region where the nucleon density is finite, 
but vanishes far outside the nucleus, 
as I am interested in a process where a
pn pair is added to or removed from a nucleus. Specifically, I choose a Woods--Saxon function
\begin{align}
    f(r)=\dfrac{1}{1+e^{(r-R)/a}}
    \label{eq:WS_form}
\end{align}
with $R = 1.27 \times A^{1/3} \,{\rm fm}$ 
and $a = 0.67 \,{\rm fm}$, 
but as shown later the main conclusion does not
depend on a detailed form of $f(r)$. 
Here, $Y_{00}$ is a spherical harmonics 
with rank 0 and both $\hat{P}^{\rm ad}$ and
$\hat{P}^{\rm rm}$ carry the angular and parity quantum numbers $0^+$. 
I describe the pn-pair vibration with spin--parity 
$J^\pi = 0^+$ and $1^+$ for $(S,T)=(0,1)$ and 
$(1,0)$, respectively.

I investigate the static polarizability 
as a measure of the pn pairing. 
One may consider the polarizability 
for the hermitian operators $\hat{P}^{\rm ad}+\hat{P}^{\rm rm}$ 
and $i(\hat{P}^{\rm ad}-\hat{P}^{\rm rm})$. 
They indeed 
correspond to the operators for 
the amplitude and phase modes, respectively~\cite{Takahashi:2023rfn}. 
Therefore, the order parameter of the pn pairing is 
$
    \langle \phi_0|\hat{P}^{\rm A}_{ST}|\phi_0 \rangle
$
with $\hat{P}^{\rm A}_{ST}=\hat{P}^{\rm{ad}}_{ST}+\hat{P}^{\rm rm}_{ST}$. 
The polarizability appropriate for the pair collectivity is then evaluated as 
\begin{align}
    \alpha_{ST}=2\sum_{n}\dfrac{|\langle n|\hat{P}^{\rm{A}}_{ST}|0\rangle|^2}{\omega_n},
\end{align}
which is an extension of the two-neutron pair polarizability in Ref.~\cite{Takahashi:2023rfn} to the case for the pn pair.

I assume in the present investigation that 
the condensation of the pn pairs ($T_z=0$) 
does not occur 
in the ground state: 
$\langle \phi_0|\hat{P}^{\rm ad}_{ST}|\phi_0\rangle =
\langle \phi_0|\hat{P}^{\rm rm}_{ST}|\phi_0\rangle=0$ for both $(S,T)=(0,1)$ and $(1,0)$. 
In this case, the pn-pair polarizability reads
\begin{align}
    \alpha_{ST}=2\sum_{n}\left(\dfrac{|\langle n|\hat{P}^{\rm{ad}}_{ST}|0\rangle|^2}{\omega_n}+\dfrac{|\langle n|\hat{P}^{\rm{rm}}_{ST}|0\rangle|^2}{\omega_n}\right).
    \label{eq:pair_pol}
\end{align}
If the system tends to break the symmetry, 
one sees the polarizability gets diverged, 
or one obtains the imaginary solution of the RPA equation.

The static polarizability depends on the form of an operator. 
The reaction mechanism also enters into the form factor. 
In envisioning the pn-transfer reactions, 
such as $(p,{}^{3}{\rm He})$ and $({}^{3}{\rm He},p)$, 
in which the isoscalar (IS) and isovector (IV) transfers are both allowed, 
what is proposed here is the ratio of the polarizabilities of the pn pair with $(S,T)=(0,1)$ and 
$(1,0)$:
\begin{align}
    \mathcal{R}_{01}:=3\dfrac{\alpha_{01}}{\alpha_{10}}.
\end{align}
A factor of three is introduced here 
to cancel out the spin degeneracy $2S+1$. 
When the $\mathcal{R}_{01}$ value is greater than unity, 
the $S=0, T=1$ pn pairing is dominant over 
the $S=1, T=0$ pn pairing. 
As in the ratio of the pn-transfer 
cross sections from the ground state of an even-even nucleus 
to the $0_1^+$ and $1_1^+$ states in an odd-odd nucleus, 
one can expect 
systematic uncertainties and, to some extent, 
kinematic conditions of the reaction cancel out~\cite{LAY2022136789}.

\subsection{Numerical Calculations}\label{Sec.numer}

As a numerical approach for the present investigation, 
I employ the nuclear energy-density functional (EDF) method. Since the details can be found in Ref.~\cite{Yoshida:2013bma}, 
here I briefly recapitulate the basic
equations relevant to the present study. 

The ground state of a mother (target) nucleus is described by solving the HFB or 
Kohn--Sham--Bogoliubov (KSB) equation~\cite{dob84}:
\begin{align}
\sum_{s^\prime}
\begin{bmatrix}
h^q_{s s^\prime}(\br)-\lambda^{q}\delta_{s s^\prime} & \tilde{h}^q_{s s^\prime}(\br) \\
\tilde{h}^q_{s s^\prime}(\br) & -h^q_{s s^\prime}(\br)+\lambda^q\delta_{s s^\prime}
\end{bmatrix}
\begin{bmatrix}
\varphi^{q}_{1,\alpha}(\br s^\prime) \\
\varphi^{q}_{2,\alpha}(\br s^\prime)
\end{bmatrix} \notag \\
= E_{\alpha}
\begin{bmatrix}
\varphi^{q}_{1,\alpha}(\br s) \\
\varphi^{q}_{2,\alpha}(\br s)
\end{bmatrix}, \label{HFB_eq}
\end{align}
where 
the single-particle and pair Hamiltonians, $h^q_{s s^\prime}(\br)$ and $\tilde{h}^q_{s s^\prime}(\br)$, 
are given by the functional derivative of the EDF with respect to the particle density and the pair density, respectively. 
An explicit expression of the Hamiltonians is found in the Appendix of Ref.~\cite{kas21}. 
The superscript $q$ denotes 
$\nu$ (neutron, $ t= 1/2$) or $\pi$ (proton, $t =-1/2$). 
The average particle number is fixed at the desired value by adjusting the chemical potential $\lambda^q$. 
When the system is in the normal phase, 
the chemical potential is defined by averaging the 
highest occupied single-particle energy 
and the lowest unoccupied single-particle energy. 
Assuming the system is axially symmetric, 
the KSB equation (\ref{HFB_eq}) is block diagonalized 
according to the quantum number $\Omega$, the $z$-component of the total angular momentum. 
As mentioned in the above, I further assumed that 
the condensation of the pn pairs ($T_z=0$) does not occur 
but that of the like-particle pairs 
($T_z=\pm 1$) can happen in the ground state. 

I solve the KSB equation in the coordinate space using cylindrical coordinates
$\boldsymbol{r}=(\varrho,z,\phi)$.
Since I assume further the reflection symmetry, only the region of $z\geq 0$ is considered. 
I use a two-dimensional lattice mesh with 
$\varrho_i=(i-1/2)h$, $z_j=(j-1)h$ ($i,j=1,2,\dots$) 
with a mesh size of
$h=0.6$ fm and 25 points for each direction. 
The qp states are truncated according to the qp 
energy cutoff at 60 MeV, and 
the qp states up to the magnetic quantum number $\Omega=31/2$
with positive and negative parities are included. 

The excited states in a neighboring odd-odd nucleus 
$| n \rangle$ are described as 
one-phonon excitations built on the ground state $|0\rangle$ of the even-even mother nucleus as 
\begin{align}
| n \rangle &= \hat{\Gamma}^\dagger_n |0 \rangle, \\
\hat{\Gamma}^\dagger_n &= \sum_{\alpha \beta}\left\{
X_{\alpha \beta}^n \hat{a}^\dagger_{\alpha}\hat{a}^\dagger_{\beta}
-Y_{\alpha \beta}^n \hat{a}_{\bar{\beta}}\hat{a}_{\bar{\alpha}}\right\},
\end{align}
where $\hat{a}^\dagger$ and $\hat{a}$ are 
the quasiparticle (qp) creation and annihilation operators that 
are defined in terms of the solutions of the KSB equation (\ref{HFB_eq}) with the Bogoliubov transformation.
The phonon states, the amplitudes $X^n, Y^n$ and the vibrational frequency $\omega_n$, 
are obtained in the quasiparticle-RPA (QRPA): the linearized time-dependent density-functional theory for superfluid systems~\cite{nak16}. 
The EDF gives the residual interactions entering into the QRPA equation. 
For the axially symmetric nuclei, the QRPA equation 
is block diagonalized according to the quantum number $K=\Omega_\alpha + \Omega_\beta$. 

In solving the QRPA equation, 
I introduce the truncation for the two-quasiparticle (2qp) configurations,
in terms of the 2qp-energy as 60 MeV. 
The calculated energy and transition strength of 
the low-lying and giant resonance states are almost converged with respect to the mesh size, the box size, and the energy cutoff~\cite{yos08}, 
and are compatible with the results obtained in different methodology~\cite{eba10,sca14}.

For the normal (particle--hole) part of the EDF $\mathcal{E}_{\rm ph}$,
I employ the Skyrme-type SGII functional~\cite{gia81}.
For the pairing energy $\mathcal{E}_{\rm pair}$, I adopt the one in Ref.~\cite{yam09}
that depends on both the IS and IV densities, 
in addition to the pair density, with the parameters given in Table~III of Ref.~\cite{yam09}. 
The same pair interaction is employed for the dynamical pairing for the $S=0$ and $S=1$ pn pairs 
in the QRPA calculation, 
while the linear term in the IV density is dropped. 
I multiply a factor $f$ to the pairing EDF for $S=1$: 
$\mathcal{E}_{\rm pair}^{S=1}=f\times \mathcal{E}_{\rm pair}^{S=0}$ 
to study the roles of the dynamic spin-triplet pairing. 
Note that the QRPA calculations including the dynamic spin-triplet pairing with 
more or less the same strength as the spin-singlet pairing $(f \sim 1)$
describe well the characteristic low-lying Gamow--Teller 
strength distributions in the light $N \simeq Z$ nuclei~\cite{fuj14,fuj15,fuj19}, and the $\beta$-decay half-lives of neutron-rich Ni isotopes~\cite{yos19}.

\section{Results and discussion}\label{Sec.Result}
\subsection{Ca isotopes}\label{Sec.Ca}

\begin{figure}[t]
    \centering
    \includegraphics[scale=0.5]{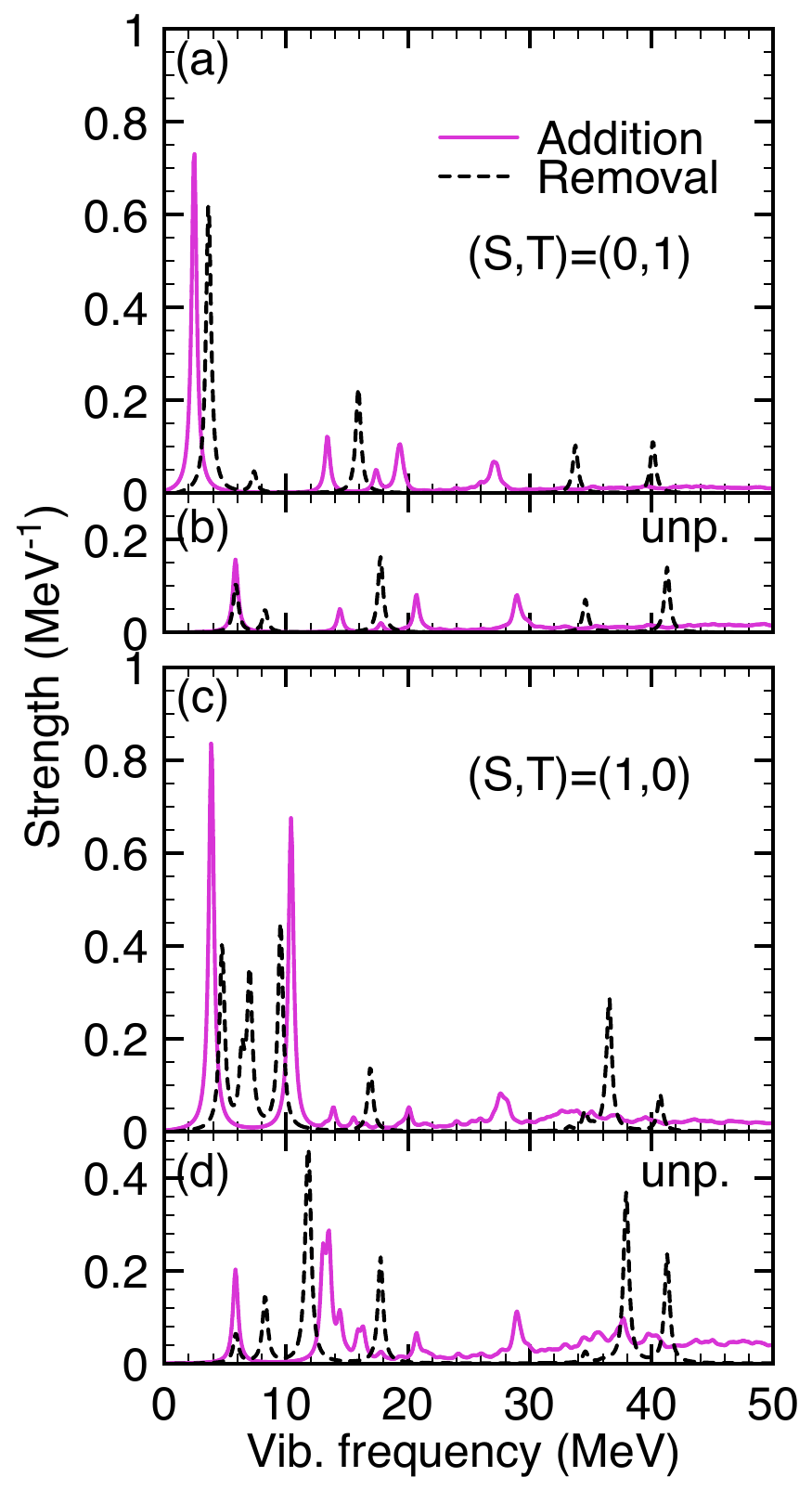}
    \caption{\label{fig:40Ca_pair_strength} 
    Calculated pair addition and removal strengths 
    in $^{40}$Ca 
    smeared by a Lorentzian function with a width parameter of 0.5 MeV. 
    (a): $S=0$ pn-pair strengths.
    (b): Unperturbed $S=0$ pn-pair strengths.
    (c): $S=1$ pn-pair strengths with $f=1.0$. 
    (d): Unperturbed $S=1$ pn-pair strengths.
}
\end{figure}

Figure~\ref{fig:40Ca_pair_strength} shows 
the calculated strength functions 
for the pair-addition and removal processes 
in $^{40}$Ca. 
A possible appearance of the collective 
states in the pn-pair transfer from ${}^{40}$Ca 
was discussed in Ref.~\cite{Yoshida:2014jga}. 
However, 
only the lowest $1^+$ state in the odd-odd neighboring nuclei, $^{42}{\rm Sc}$ and $^{38}{\rm K}$, 
was investigated as a collective state of 
the $S=1$ pn pairing. 

As is observed in the figure, 
not only the lowest state but also the higher-lying states acquire appreciable strengths 
for both the $S=0$ and $S=1$ pn-pair transfers. 
The occurrence of the higher-lying states is due to 
the shell structure: 
The second $0^+$ state seen in the pn-pair addition 
is mainly generated by the $\nu p_{3/2}\otimes\nu p_{3/2}$ configuration with an admixture of 
$\nu p_{1/2}\otimes\nu p_{1/2}$; 
while the second $1^+$ state is constructed by the 
$\nu p_{3/2}\otimes\nu p_{1/2}$ and 
$\nu p_{1/2}\otimes\nu p_{3/2}$ configurations additionally. 
One can see there are several states even above 20 MeV.  
These states may also be regarded as
pair-vibrational states as the excitation energies 
are lowered 
and the pair-transfer strengths are enhanced by the RPA correlation, 
as compared with the unperturbed strength functions 
depicted in Figs.~\ref{fig:40Ca_pair_strength}(b) and \ref{fig:40Ca_pair_strength}(d). 
Although high-lying pair vibration is often called the giant pair vibration~\cite{BROGLIA1977129,cap15, ass19, cav19}, 
I adopt in the present paper, 
following Ref.~\cite{Takahashi:2023rfn}, 
the more inclusive term ``high-lying pair vibrations'' 
for these peaks
since the strength distribution does not 
form a single resonance peak 
but rather multiple peaks in a wide energy interval.

\begin{figure}[t]
    \centering
    \includegraphics[scale=0.45]{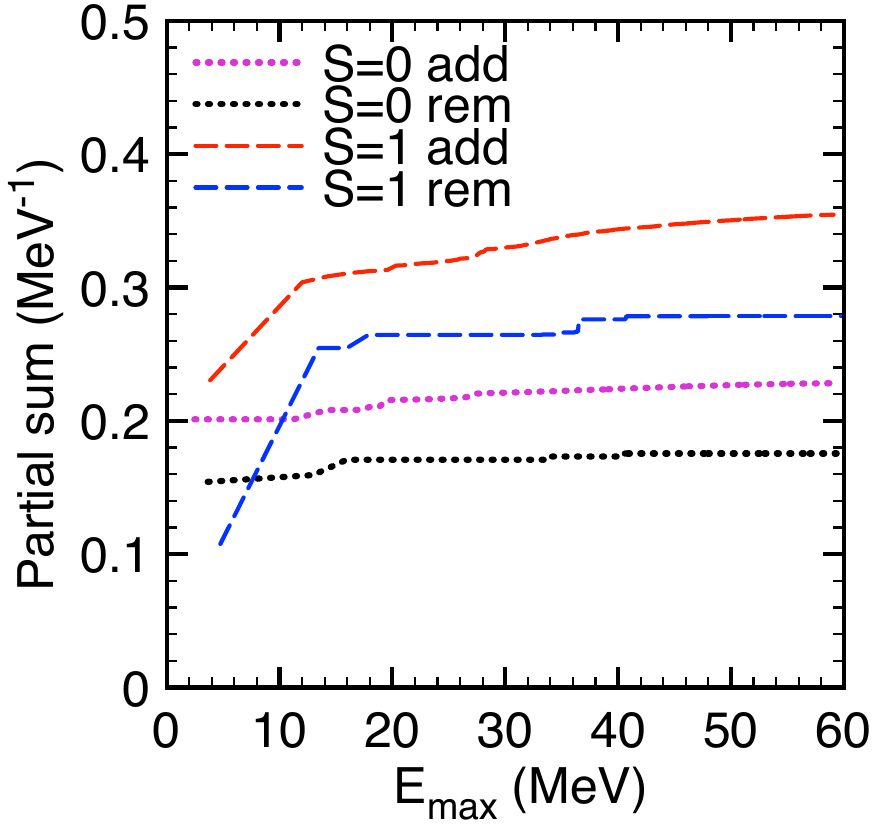}
    \caption{\label{fig:40Ca_sum} 
    The inversely-energy-weighted sum of the pn-pair addition and removal strength functions for $^{40}$Ca, 
    as a function of the maximum energy $E_{\rm max}$ of the sum: $\omega_n<E_{\rm max}$.
}
\end{figure}

The primal motivation to introduce 
the pair polarizability in the present study is to 
relate the pair vibrations distributed in a broad energy 
region with the macroscopic picture as an amplitude oscillation of the order parameter. 
The polarizability is defined by the inversely-energy-weight sum of the strength functions, as given in Eq.~(\ref{eq:pair_pol}). 
The low-lying pair vibrational states 
make a substantial contribution to polarizability. 
The high-lying pair vibrations can additionally 
play a role. 
Figure~\ref{fig:40Ca_sum} shows a running sum 
in Eq.~(\ref{eq:pair_pol}) up to $E_{\rm max}$, 
where the upper bound $E_{\rm max}$ is varied 
for $^{40}$Ca with $f=1.0$. 
The low-lying states below 15 MeV are dominant 
for the polarizability, and 
more than 90\% of the polarizability 
is given by the states up to 25 MeV.
This is similar to the finding in Ref.~\cite{Takahashi:2023rfn}.

\begin{figure}[t]
    \centering
    \includegraphics[scale=0.5]{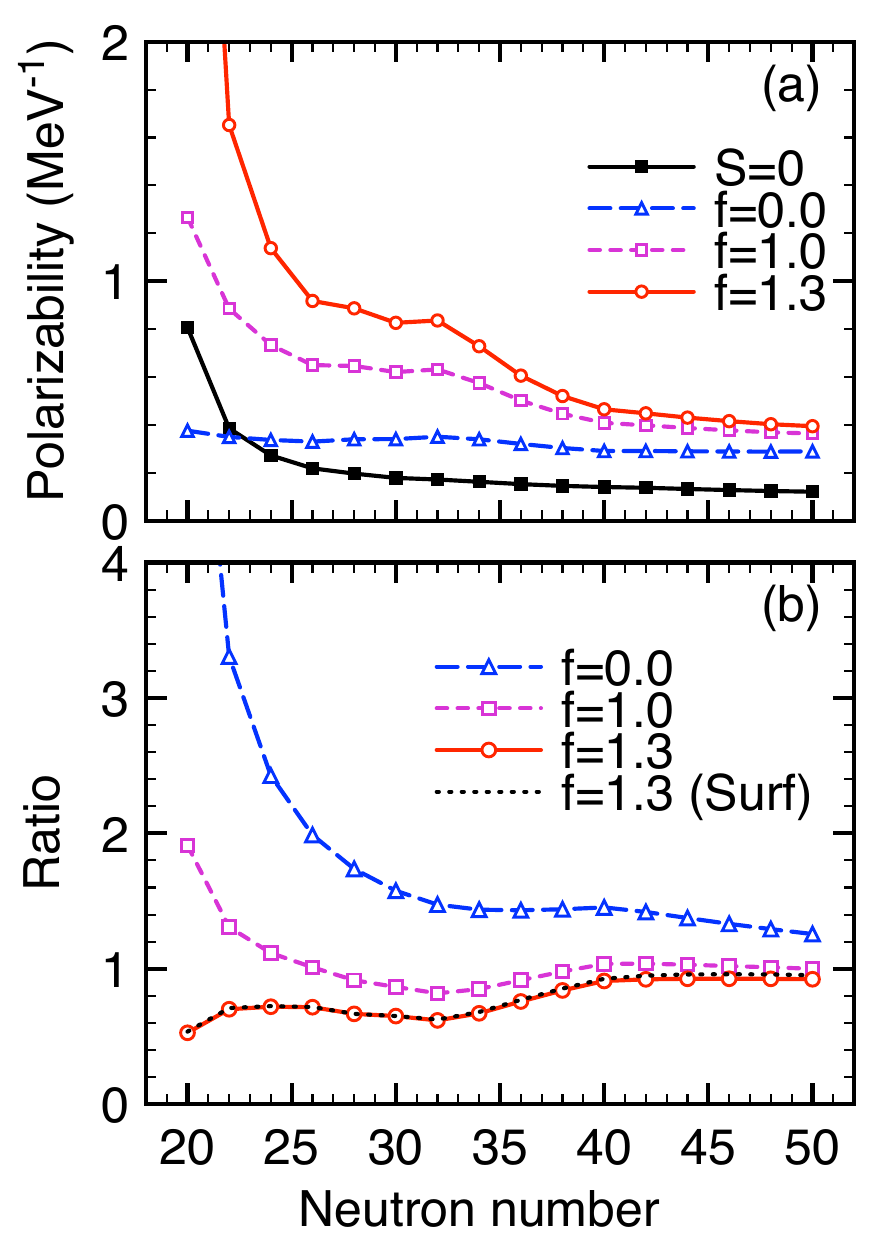}
    \caption{\label{fig:Ca_pair_pol} 
    (a): Polarizalibities for the pn-pair transfers with $S=0$ and $S=1$ for the Ca isotopes. For those with $S=1$, 
    the results obtained by changing $f$ are also shown.
    (b): The ratio $\mathcal{R}_{01}$ calculated by changing $f$. The results obtained by using the surface-type form factor are also shown for the case of $f=1.3$ (Surf).
}
\end{figure}

Figure~\ref{fig:Ca_pair_pol}(a) shows the 
polarizabilities for 
pn-pair transfers with $(S,T)=(0,1)$ and $(1,0)$.
The $S=1$ pn-pair polarizability has a peak 
at $N=Z$ and drops 
with an increase in the neutron number. 
This is what one would expect: 
the pn-pair correlation is strongest at $N=Z$ 
and gets weaker with the neutron excess. 
Without the $S=1$ pn-pair interaction ($f=0$), 
the $S=1$ pn-pair polarizability keeps constant 
at a low value. 
When considering the pn-pair interaction 
in the $S=1$ channel with $f>0$, 
the polarizability shows maximal at $N=Z$ 
and decreases
with increasing the neutron number.  
However, one sees the persistence around $N=30$. 
Beyond ${}^{48}$Ca, 
the $\nu f_{5/2}$ orbital is located near the 
Fermi level, and then the $S=1$ pairs of 
the $\pi f_{7/2}\otimes \nu f_{5/2}$ configuration 
are created with low energy.

It would not be easy to obtain 
the absolute values of the pair polarizability 
since the reaction condition is involved.
Instead, 
the isotopic dependence of 
the pair polarizability of $S=0$ and $S=1$ 
provides us with information on pair collectivity, 
as the shell effect uniquely appears in 
the $S=1$ pn-pair transfer. 
Alternatively, by taking the ratio of 
the pair polarizabilities of $S=0$ and $S=1$, 
one can gain insight into the pairing collectivity. 
The ratio $\mathcal{R}_{01}$ for the 
Ca isotopes are shown in 
Fig.~\ref{fig:Ca_pair_pol}(b). 
Without the pn-pair interaction 
in the $S=1$ channel, 
it is evident that the $S=0$ pairing is dominant 
in $N \approx Z$ isotopes: 
the $\mathcal{R}_{01}$ value is far above the unity. 
Even with a reasonable strength ($f=1.0$) 
of the $S=1$ pn pair interaction, 
the $S=0$ pairing is stronger than 
the $S=1$ pairing around $N=Z$. 
However, 
the $S=0$ and $S=1$ pair correlations vie in collectivity 
around ${}^{50}$Ca.

In the calculation with $f=1.3$, 
the lowest $1^+$ state is lower than the $0^+$ state by 0.6 (0.2) MeV for the pn-pair addition (removal) transfer, 
and the transition strength to the $1^+$ state is 
strongly enhanced for $^{40}$Ca. 
It seems that the system is close to the critical point 
of the $S=1$ pn-pair condensation; 
the $\mathcal{R}_{01}$ value is about 0.5. 
In reality, however, 
the $S=1$ pn-pairing is not so strong 
because the $1^+$ state is higher in energy than 
the $0^+$ state in $^{42}$Sc and $^{38}$K. 

The method to evaluate the pn-pair collectivity, 
the $\mathcal{R}_{01}$ value, 
from the response should not
depend on details of the definitions of the pn-pair transfer operators, 
$\hat{P}^{\rm ad}, \hat{P}^{\rm rm}, \hat{P}^{\rm A}$.  
I have examined another choice of the form factor $f(r)$ appearing in Eqs.~(\ref{eq:pair_ad}) and (\ref{eq:pair_rem}) by 
using $f(r)=(d/dr)[1+e^{(r-R)/a}]^{-1}$, 
which has a weight on the surface area of the nucleus. 
The results obtained by using this surface-type form factor are shown in Fig.~\ref{fig:Ca_pair_pol}(b) 
for the case of $f=1.3$, in which the amplitude of 
the transition density is high 
and thus the results may be sensitive to 
the spatial dependence of the operator. 
As one can see in the figure, 
the $\mathcal{R}_{01}$ value 
obtained is almost the same as that obtained 
by the volume-type form factor. 
I can conclude that the $\mathcal{R}_{01}$ value 
does not depend 
on the choice of the form factor. 

\subsection{Ni and Sn isotopes}\label{Sec.NiSn}

\begin{figure}[t]
    \centering
    \includegraphics[scale=0.29]{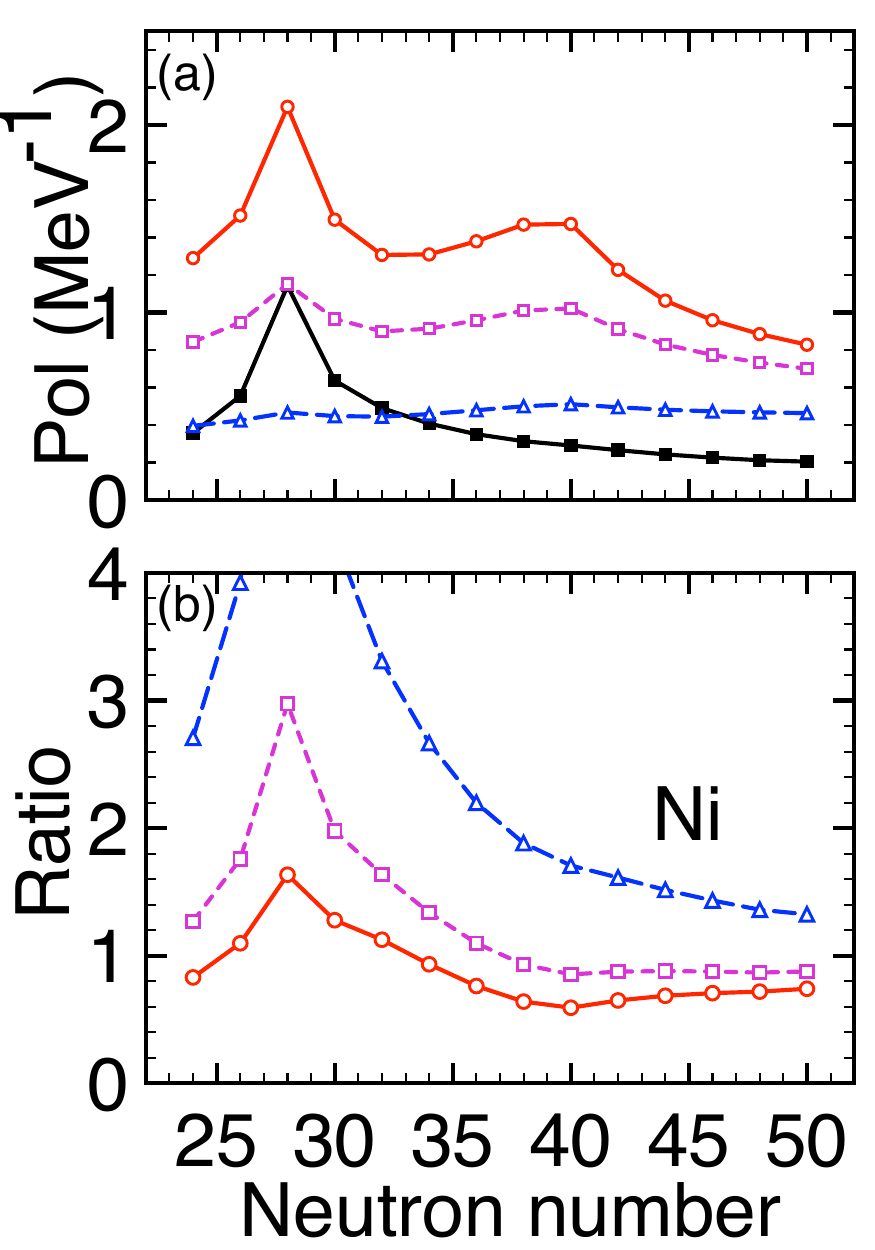}
    \includegraphics[scale=0.29]{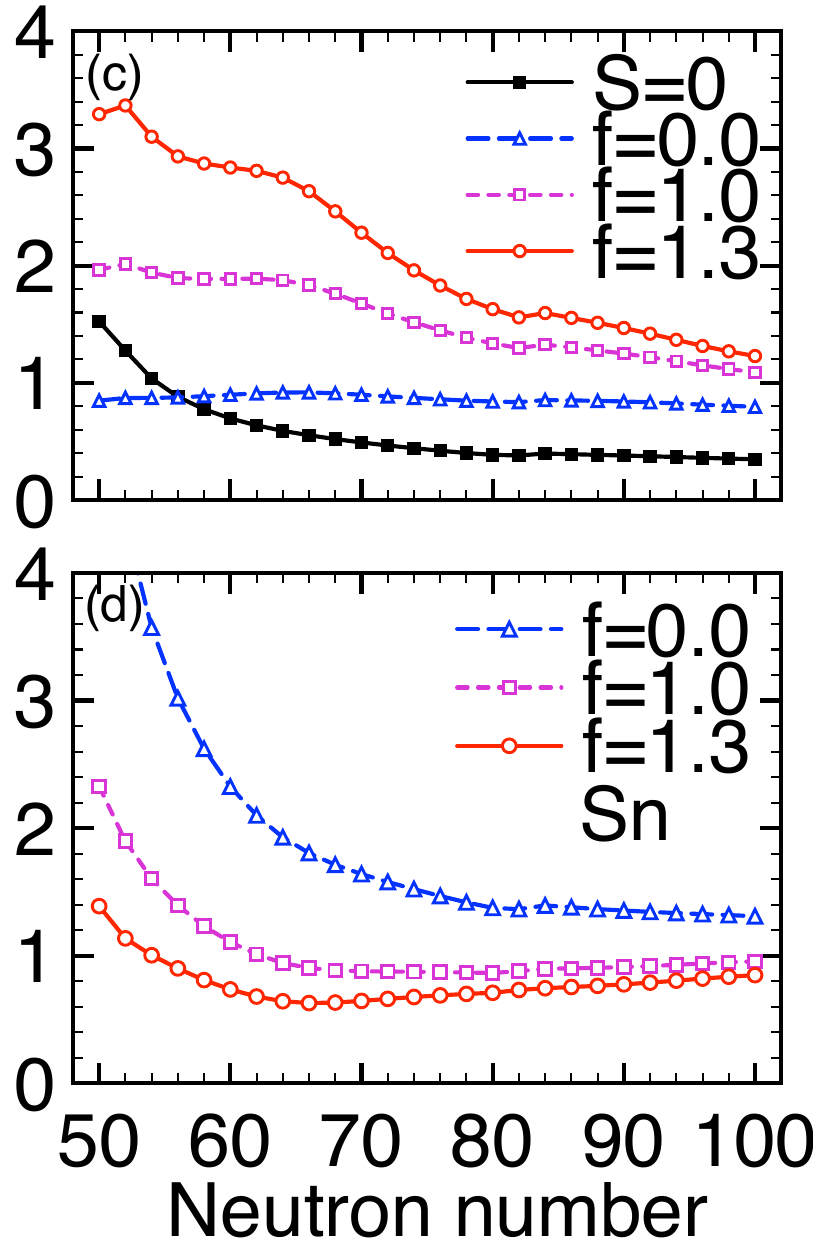}
    \caption{\label{fig:Ni_Sn_pair_pol} 
    As Fig.~\ref{fig:Ca_pair_pol} but for the Ni and Sn isotopes. 
}
\end{figure}

The shell effect in the $S=1$ pn-pairing is further 
investigated for the Ni and Sn isotopes. 
The calculated pn-pair polarizabilities $\alpha_{01}, \alpha_{10}$ and the ratio $\mathcal{R}_{01}$
are shown in Fig.~\ref{fig:Ni_Sn_pair_pol}. 
As in the case of the Ca isotopes, 
the pn-pair polarizabilities are large at $N \approx Z$ 
for both $S=0$ and $S=1$ with inclusion 
of the spin-triplet pair interaction ($f>0$).
However, the $S=1$ pn-pair polarizability $\alpha_{10}$
increases toward $N=40$ in the Ni isotopes 
and keeps almost constant 
up to $N=64$ in the Sn isotopes. 
This isotopic dependence is clearly seen in the ratio $\mathcal{R}_{01}$: 
one sees a minimum at $N=40$ and 64 
in the Ni and Sn isotope, respectively; 
the $S=1$ pn-pair correlation is stronger 
than the $S=0$ one.

\begin{figure}[t]
    \centering
    \includegraphics[scale=0.45]{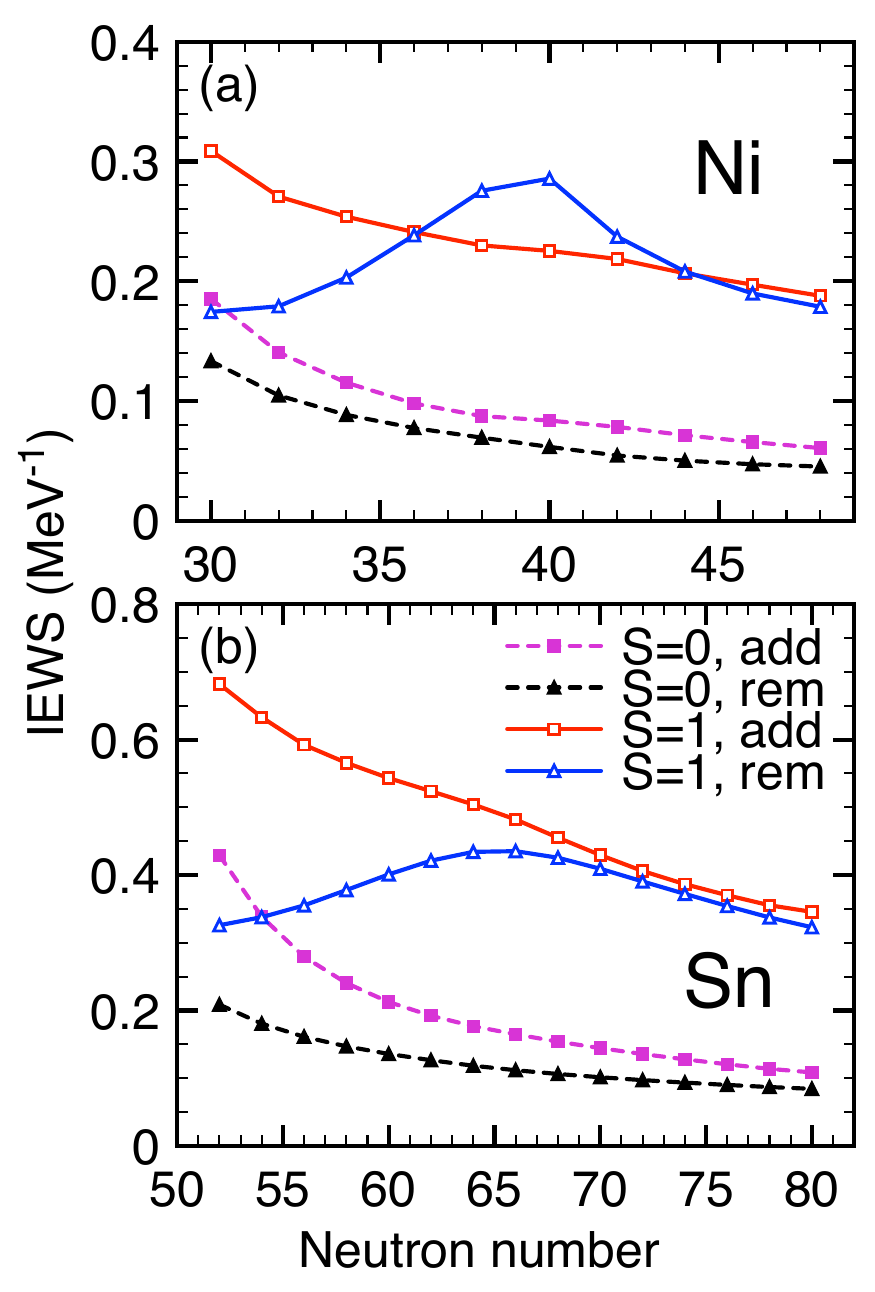}
    \caption{\label{fig:Ni_Sn_IEWS} 
    IEWS of the pn-pair 
    addition and removal strengths in (a) Ni 
    and (b) Sn isotopes, 
    calculated by using $f=1.0$ for the spin-triplet 
    pair interaction.
}
\end{figure}

The shell effect at $N=40$ and 64 is not 
due to the appearance of the $S=1$ pn-pair addition 
mode, as seen in the Ca isotopes, 
but due to that of the $S=1$ pn-pair removal 
mode in low energy. 
Figure~\ref{fig:Ni_Sn_IEWS} shows the IEWS of the 
pn-pair addition and removal strengths. 
The IEWS for the $S=1$ pn-pair removal 
increases toward $N=40$ and 64 in the Ni and Sn 
isotopes, respectively, while the IEWSs for 
both the $S=0$ pn-pair addition and removal 
as well as the $S=1$ pn-pair addition 
decrease monotonically as stepping away from $N=Z$. 
It is noted that the IWSEs for the 
$S=1$ pn-pair addition and removal 
keep almost constant as functions of the neutron number 
when the spin-triplet pair interaction is discarded, 
as expected from the calculated pair polarizability $\alpha_{10}$ with $f=0$; 
the enhancement in the IWSE for the 
$S=1$ pn-pair removal is much more significant 
in the calculations with $f>1$.

A decisive role in 
the shell effect uniquely appearing in the $S=1$ pn-pairing in the medium-heavy nuclei 
such as in the Ni and Sn isotopes 
is played by the occupation of the 
intruder proton $j_>$ orbital. 
Indeed, this diminishes the $S=1$ pn-pairing 
in the $N \approx Z$ nuclei beyond $^{40}$Ca: 
only either the $\pi j_> \otimes \nu j_>$ or 
$\pi j_< \otimes \nu j_<$ configuration 
is devoted to the pairing. 
With an increase in the neutron number from $N=Z$, 
the Fermi level of neutrons approaches  
the $j_<$ orbital. 
As neutrons start to occupy the $j_<$ orbital, 
the pair of 
the $\pi j_> \otimes \nu j_<$ configuration, 
which is exclusively in $S=1$, appears to 
contribute to the pairing collectivity. 
In the present case, 
the pair of the $\pi f_{7/2}\otimes \nu f_{5/2}$ 
and $\pi g_{9/2} \otimes \nu g_{7/2}$ configuration 
gives a significant contribution to 
the $S=1$ pairing in the Ni and Sn isotopes, 
respectively. 

\section{Summary}\label{Sec.Sum}

Proton-neutron (pn) pair correlations have been 
investigated in the Ca, Ni, and Sn isotopes 
as examples of medium-heavy nuclei. 
The collectivity was quantified 
by introducing the polarizability 
for the response to pn-pair transfers. 
The polarizabilities were evaluated 
in the framework of the nuclear energy-density functional 
method. 

The spin-singlet pn-pair correlation is the strongest 
at $N=Z$ 
and decreases monotonically as the neutron number increases. 
In contrast, the spin-triplet pn-pair correlation 
is shown to depend non-monotonically on the neutron number 
and can be enhanced in cases where 
the pn-pair removal modes 
involving the $\pi j_> \otimes \nu j_<$ 
configuration occur at low energy. 
Therefore, the relative strength of spin-triplet 
and spin-singlet pn-pair correlations becomes large 
in the neutron-rich isotopes 
when neutrons occupy the $j_<$ orbital.

\section*{Acknowledgments}
The author acknowledges the invaluable contributions of members of the PHANES Collaboration, particularly M.~Dozono, M.~Matsuo, S.~Ota, and S.~Shimoura, 
for their insightful discussions. 
This work was supported by
JSPS KAKENHI (Grant No. JP23H05434), 
and JST ERATO (Grant No. JPMJER2304).
The numerical calculations were performed on the computing facilities  
at the Yukawa Institute for Theoretical Physics, Kyoto University, and 
at the Research Center for Nuclear Physics, Osaka University.  
This work also used computational resources of SQUID 
provided by the D3 Center, Osaka University 
through the HPCI System Research Project 
(Project ID: hp230085).

\end{document}